\documentclass{article}

\usepackage{PRIMEarxiv}
\usepackage{setspace}
\usepackage[utf8]{inputenc} 
\usepackage{amsmath}        
\DeclareMathOperator*{\argmax}{arg\,max}
\usepackage{xcolor}         
\usepackage[T1]{fontenc}    
\usepackage{hyperref}       
\usepackage{url}            
\usepackage{booktabs}       
\usepackage{amsfonts}       
\usepackage{nicefrac}       
\usepackage{microtype}      
\usepackage{lipsum}
\usepackage{cite}
\usepackage{fancyhdr}       
\usepackage{graphicx}       
\usepackage{xcolor}
\usepackage[utf8]{inputenc}
\usepackage[english]{babel}
\usepackage{lineno}

\graphicspath{{media/}}     

\title{Simultaneous quantification and changepoint detection of point source gas emissions using recursive Bayesian inference}

\author{  \href{https://orcid.org/0000-0003-2167-8614}{\includegraphics[scale=0.06]{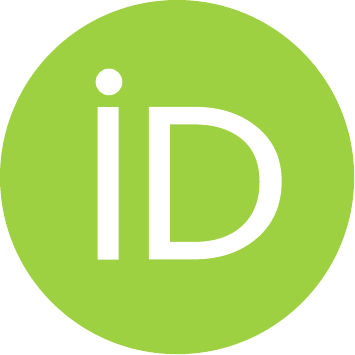}\hspace{1mm}}
 Amir Montazeri \\
  Sibley School of Mechanical and Aerospace Engineering\\
  Cornell University\\
  Ithaca, NY, USA\\
  \href{mailto:am2774@cornell.edu}{\texttt{am2774@cornell.edu}}\\
   \\ 
   \And
 \href{https://orcid.org/0000-0002-7123-3835}{\includegraphics[scale=0.06]{orcid.pdf}\hspace{1mm}} Xiaochi Zhou \\
  School of Civil and Environmental Engineering \\
  Cornell University\\
  Ithaca, NY, USA \\
  \href{mailto:xiaochi.zhou@arb.ca.gov}{\texttt{xiaochi.zhou@arb.ca.gov}} \\
  \And
 John D. Albertson\thanks{Corresponding author.} \\
  School of Civil and Environmental Engineering \\
  Cornell University\\
  Ithaca, NY, USA \\
  \href{mailto:albertson@cornell.edu}{\texttt{albertson@cornell.edu}}\\

  \\
}

\begin{document}
\maketitle

\begin{abstract}
Recent findings suggest that abnormal operating conditions of equipment in the oil and gas supply chain represent a large fraction of anthropogenic methane emissions. Thus, effective mitigation of emissions necessitates rapid identification and repair of sources caused by faulty equipment. In addition to advances in sensing technology that allow for more frequent surveillance, prompt and cost-effective identification of sources requires computational frameworks that provide automatic fault detection.  Here, we present a changepoint detection algorithm based on a recursive Bayesian scheme that allows for simultaneous emission rate estimation and fault detection. The proposed algorithm is tested on a series of near-field controlled release mobile experiments, with promising results demonstrating successful detection (>90\% success rate) of changes in the leak rate when the emission rate is tripled after an abrupt change. Moreover, we show that the statistics of the measurements, such as the coefficient of variation and range are good predictors of the performance of the algorithm. Finally, we describe how this methodology can be easily adapted to suit time-averaged concentration data measured by stationary sensors, thus showcasing its flexibility. 
\end{abstract}


\section{Introduction}
Methane is a potent greenhouse gas (GHG) with a global warming potential (GWP) that is approximately 84 and 28 times greater than carbon dioxide (CO\textsubscript{2}) on 20 and 100 year time scales, respectively \cite{stocker_climate_2014}. Methane emissions from the oil and gas industry are among the largest anthropogenic sources of methane in the United States, accounting for approximately 30\% of total emissions in 2019 \cite{usepa_2021}. Recent studies have found that emissions from almost all subsectors of the oil and gas supply chain demonstrate a “fat-tail” distribution, such that a relatively small number of large emitters are responsible a large fraction of the emissions \cite{petron_new_2014,brandt_methane_2014, frankenberg_airborne_2016,ravikumar_good_2018,zavala-araiza_reconciling_2015,zavala-araiza_toward_2015,zhou_mobile_2021}. While the presence of these large emitters is concerning, it offers the potential of expedient reduction in GHG emissions and costs if the largest emitters are rapidly identified and repaired. 

Measurements of methane emissions are often classified as either top-down or bottom-up \cite{alvarez_assessment_2018}. Top-down studies rely on ambient methane measurements using aircraft, satellites, or tower networks to estimate aggregate emissions from all contributing sources across large geographies \cite{alvarez_assessment_2018}. On the other hand, bottom-up methods aggregate and extrapolate emissions from individual pieces of equipment, operations, or facilities, using measurements made directly at the emission point or, in the case of facilities, directly downwind \cite{alvarez_assessment_2018, rella_measuring_2015,omara_methane_2016,robertson_variation_2017,brandt_methane_2016}. Recent integrated research efforts have found that while emission estimates from facility-based bottom-up approaches and top-down approaches are in agreement, these estimates are significantly higher than component-based estimates (i.e., when emissions are extrapolated from individual pieces of equipment) [3-4]. Detailed investigation of the discrepancy between component-based aggregates and estimates from other approaches (i.e., facility-based bottom-up and top-down methods) suggests that component-based methods miss high emissions caused by abnormal operating conditions (e.g., malfunctions). Such abnormal conditions are the defining attribute of large emitters that contribute the majority of emissions in the oil and gas sector. Therefore, prompt identification of abnormal process conditions (i.e., fault detection) can lead to substantial reductions in emission and costs for operators. 

The abnormal operating conditions observed in the largest emitters are spatially and temporally variable \cite{vaughn_temporal_2018, zavala-araiza_super-emitters_2017, duren_californias_2019,brandt_methane_2014}. For example, emissions can significantly increase for a production site due to malfunctions at a certain point in time. Hence, emission reduction requires monitoring approaches that enable efficient and timely responses to the appearance of abnormal process conditions. Continuous monitoring offers the capability to rapidly detect faulty behavior that is necessary for reducing emissions from the largest emitters. Moreover, recent efforts to develop innovative technologies and algorithms to mitigate methane emissions have also highlighted the advantages of continuous monitoring over “snapshot-in-time” approaches in rapid identification of large emission sources \cite{coburn_regional_2018, arpa2015}. 

While continuous monitoring of oil and gas facilities are not commonplace yet, changes in the near future are likely due to the following: 1) Advancements in sensor technology and wireless communications allow for continuous measurements to be made and stored in the cloud \cite{coburn_regional_2018}, and 2) monitoring mandates at the state and federal level, such as the U.S. Environmental Protection Agency (USEPA) fence-line monitoring program for early detection of benzene emissions \cite{usepa_2020} and continuous monitoring of air quality during pre-production and early-production of drilling operations producing gas and liquid hydrocarbons required by the Colorado Department of Public Health and environment \cite{CDPHE_2021}. Meanwhile, innovations in atmospheric inversion modeling are required to enable automatic emission estimation and fault detection using continuous measurements. It is worth noting that efficient and rapid fault detection can serve as an incentive for operators to employ continuous surveillance systems, as it can lead to significant reductions in unwanted emissions and costs. 

Here, we propose a recursive Bayesian inference model that utilizes measurements from continuous surveillance systems (e.g., network of fixed sensors, and mobile sensors on unmanned vehicles) for simultaneous estimation and changepoint (fault) detection in point-source emission rates. Note that we use changepoint or fault to refer to a sudden increase in point-source emission rate as expected under abnormal operating conditions described earlier. The Bayesian inference model is useful in this context for multiple reasons: 1) It is equipped to deal with noisy data which in this case are caused by the stochastic nature of turbulence that drives the emitted gas and other measurement uncertainties, 2) it permits the determination of the uncertainty in estimated emission rates \cite{yee_bayesian_2007} and 3) it allows for “online” changepoint detection and emission estimates, i.e., the emission rate estimates and probability of detecting changes in emission rates are updated with every new measurement that arrives incrementally \cite{adams_bayesian_2007}.

In this chapter, we first introduce an instantaneous plume dispersion formulation that is used to guide our Bayesian analysis. Then, the mathematical framework of the Bayesian inference and its application in point-source estimation and changepoint detection is described. Next, we apply the proposed Bayesian framework to near-field (with source-to-sensor distances $\leq$ 30m) mobile measurements of a controlled point-source emission and evaluate its performance in changepoint detection. Although the Bayesian framework in this chapter is tailored towards mobile measurements, it can easily be adapted for continuous measurements made by networks of fixed sensors. 

\section{Theory}
\label{sec:Theory}
Point-source characterization approaches often rely on time-averaged measurements. However, our analysis of the mobile sensor data requires a formulation for the instantaneous plume that is adapted from \cite{zhou_mobile_2019} and described in section \ref{sec:Transport}. The presented formulation is applicable to passive scalars, which are diffusive contaminants in low concentrations such that they have no dynamical effect on the motion of the surrounding flowing fluid \cite{warhaft_passive_2000}. In subsequent sections, the following assumptions are made: 1) The emission rate from the point source is constant until an abrupt change causes the leak rate to increase to a new constant value. 2) The emission rates before and after the changepoint, and therefore mass concentrations before and after the changepoint are independent.

\subsection{Instantaneous view of plume transport}
\label{sec:Transport}
\begin{figure}
\centering
  \includegraphics[scale=0.9]{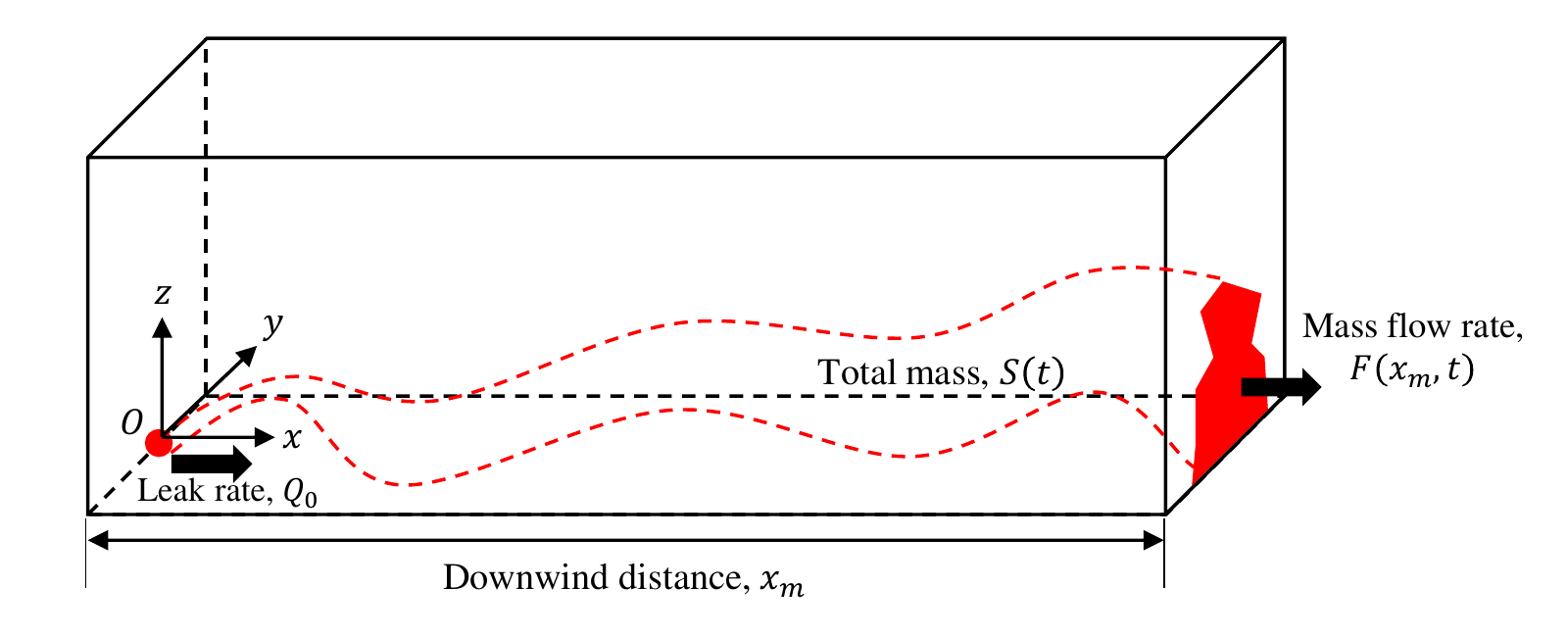}
  \caption{A control volume containing a point emission source (located at the origin, $O$) with a mass flow rate of $Q_0$, and a cross-plane view of the mass flow rate at downwind distance, $x_m$. }
  \label{fgr_CV}
\end{figure}

Consider a steady-state point source located at the origin $O$ of a local coordinate system (Figure \ref{fgr_CV}). The wind velocity components in $x, y,$ and $z$ directions are defined as $u, v,$ and $w$, respectively. We define a control volume starting at the origin $O$ to a downwind vertical plane located at $x_m$, extending from $y_{min}$ to $y_{max}$ laterally, and from $z_{min}$ to $z_{max}$ vertically, such that the control volume encompasses the entire plume upwind of the mobile sensor. For this control volume, the application of conservation of mass yields the following expression for the emission rate (mass per time), $Q_0$:
\begin{equation}
\label{eq:Q}
    Q_0 = F(x_m,t) + \frac{dS(t)}{dt},
\end{equation}
where $S(t)$ is total mass of the emitted gas in the control volume, $t$ is time, and $F(x_m,t)$ is the mass flow rate exiting the control volume through the vertical plane at $x_m$. The control volume is defined such that no mass exits anywhere other than the vertical plane at $x_m$, therefore the mass flow rate can be expressed as
\begin{equation}
\label{eq:F}
    F(x_m,t) = \int_{z_{min}}^{z_{max}} \int_{y_{min}}^{y_{max}} c(x_m,y,z,t)u(x_m,y,z,t)dydz,
\end{equation}
where $c$ is the mass concentration of the passive scalar. It is worth noting that in the atmospheric boundary layer, large Reynolds numbers are typically observed and the flow is highly turbulent, therefore, molecular diffusion is ignored \cite{stull_introduction_1988}.

It is useful to define a normalized distribution of the mass concentration, labeled $D$, and a plume-weighted advection velocity, labeled $u_e$, at the exit plane of the control volume as
\begin{equation}
\label{eq:D}
    D(x_m,y,z,t) = \frac{c(x_m,y,z,t)}{\int_{z_{min}}^{z_{max}} \int_{y_{min}}^{y_{max}}c(x_m,y,z,t)dydz}
\end{equation}
\begin{equation}
\label{eq:u_e}
    u_e(x_m,t) = \int_{z_{min}}^{z_{max}} \int_{y_{min}}^{y_{max}} D(x_m,y,z,t)u(x_m,y,z,t)dydz,
\end{equation}
We can rewrite $u_e(x_m,t)$ by substituting equation (\ref{eq:D}) into (\ref{eq:u_e}) and applying equation (\ref{eq:F}) as: 

\begin{align}
\label{eq:u_e2}
    u_e(x_m,t) &= \frac{\int_{z_{min}}^{z_{max}} \int_{y_{min}}^{y_{max}}c(x_m,y,z,t)u(x_m,y,z,t)dydz}{\int_{z_{min}}^{z_{max}} \int_{y_{min}}^{y_{max}}c(x_m,y,z,t)dydz} \nonumber \\ 
    &= \frac{F(x_m,t)}{\int_{z_{min}}^{z_{max}} \int_{y_{min}}^{y_{max}}c(x_m,y,z,t)dydz}.   
\end{align}
Equation (\ref{eq:u_e2}) can then be used in conjunction with equation (\ref{eq:Q}) to relate the mass concentration trajectory when traversing the plume, $c(x_m,y,z,t)$, to the other relevant variables as
\begin{equation}
\label{eq:c1}
    c(x_m,y,z,t) = \frac{Q_0 - dS(t)/dt}{u_e(x_m,t)}D(x_m,y,z,t).
\end{equation}
In practice, $u_e(x_m,t)$ can be approximated using nearby meteorological measurements. The vertical scaling of the wind profile based on the Monin-Obukhov similarity theory (MOST) can then be applied to adjust these meteorological measurements by height difference as needed \cite{obukhov_turbulence_1971}, following the methodology previously described in \cite{zhou_mobile_2019}. Accordingly, $u_e(x_m,t)$ is replaced by $u_e^M(x_m,t)\delta_u(x_m,t)$, where $\delta_u(x_m,t)$ accounts for the ratio between the actual and approximated plume-weighted advection velocities, $u_e(x_m,t)$ and $u_e^M(x_m,t)$, respectively. The superscript $M$ is used to highlight model estimated quantities. Similarly, we introduce $\delta_S(x_m,t)= 1-\frac{dS(t)/dt}{Q_0}$ to represent the non-steadiness in the total mass stored in the control volume, normalized by $Q_0$. Therefore, equation \ref{eq:c1} can be expressed as follows
\begin{equation}
\label{eq:c2}
    c(x_m,y,z,t)=\frac{Q_0}{u_e^M(x_m,t)} \left(\frac{\delta_S(x_m,t)}{\delta_u(x_m,t)}\right)D(x_m,y,z,t),
\end{equation}
which describes an instantaneous view of plume transport while having the same underlying form as commonly used models based on an ensemble-averaged view \cite{horst_footprint_1992}. The key differences between this instantaneous view and common ensemble-averaged models are the time dependence of the distribution $D$ that represents the stochastic nature of the turbulent plume, and the presence of $\delta_u(x_m,t)$ and $\delta_S(x_m,t)$ that accounts for non-stationarity in wind speed and mass storage in the control volume.

$D(x_m,y,z,t)$ is a random variable that captures the plume movement in time and the lateral and vertical directions as it responds to the instantaneous turbulent velocity components in these directions. It is therefore expected that $D(x_m,y,z,t)$ scales with the standard deviations of the velocity components in the $y$ and $z$ directions ($\sigma_v$ and $\sigma_w$, respectively). It is well-understood that local scaling approaches based on MOST often provide an acceptable description of $\sigma_w$ \cite{kaimal_atmospheric_1994}. However, $\sigma_v$ is affected by random large scale motions in the atmosphere that cannot be described accurately by local scaling laws \cite{lumley_structure_1964}. Therefore, a greater degree of randomness is expected in $y$ than in $z$ directions for $D(x_m,y,z,t)$. 

This observation highlights the benefits of integrating both sides of equation (\ref{eq:c2}) with respect to $y$ (i.e. across the plume), since the uncertainty associated with the lateral plume dispersion can be effectively removed \cite{albertson_mobile_2016}:
\begin{equation}
\label{eq:cy1}
    c^y(x_m,z,t) = \frac{Q_0}{u_e^M(x_m,t)} \left(\frac{\delta_S(x_m,t)}{\delta_u(x_m,t)}\right)D_z(x_m,z,t)
\end{equation}
where $c^y(x_m,z,t) =  \int_{y_{min}}^{y_{max}} c(x_m,y,z,t) dy$ is the cross-plume integrated concentration, and the reflection of the vertical profile of mass conservation at $x_m$ is described by  $D_z =  \int_{y_{min}}^{y_{max}} D(x_m,y,z,t) dy$. Note that $D_z(x_m,z,t)$ is a random variable that is mainly driven by the stochastic nature of the vertical transport dynamics in the turbulent flow. In field applications, the sensor path is typically constrained by adjacent roadways, which are not always perpendicular to the wind direction. When the road segments are at a significant angle to the wind direction, $c^y(x_m,z,t)$ can be estimated by numerical integration of the mass concentration along the path as \cite{albertson_mobile_2016}:
\begin{equation}
    c^y(x_m,z,t) = \sum_{y_{min}}^{y_{max}} c(x_m,y,z,t) \Delta t V \sin(\theta_r)
\end{equation}
where $\Delta t$ is the sensor acquisition time step, $V$ is the vehicle velocity and $theta_r$ is the acute angle between the road segment and the wind direction.

We can account for all the stochasticity in $c^y(x_m,z,t)$ by introducing a fluctuating variable. To this end, we define $D_{z,e} = \frac{\delta_S(x_m,t)}{\delta_u(x_m,t)} D_z(x_m,z,t)$ to  include the stochastic nature of $\delta_S(x_m,t), \delta_u(x_m,t)$ and $D_z(x_m,z,t)$. This new fluctuating variable is helpful in empirical analysis of the cross-plume integrated mass concentration and can be used to rewrite equation (\ref{eq:cy1}) as follows:
\begin{equation}
\label{eq:cy2}
    c^y(x_m,z,t) = \frac{Q_0}{u_e^M(x_m,t)} D_{z,e}(x_m,z,t).
\end{equation}

Equation (\ref{eq:cy2}) can be used in a forward manner to estimate the downwind cross-plume integrated mass concentration $c^y(x_m,z,t)$ for a given emission rate $Q_0$, as well as in the inverse problem of inferring $Q_0$ given downwind measurements of $c^y(x_m,z,t)$. Most dispersion models only offer an approximation of the ensemble-averaged $D_{z,e}(x_m,z,t)$, therefore, we apply Bayesian inference to account for the fluctuation of the instantaneous $D_{z,e}(x_m,z,t)$ from its ensemble averaged as detailed in section \ref{sec:Bayesian}. For simplicity of notation the independent variables, $x_m,z,$ and $t$ will be dropped hereafter. 

\subsection{Bayesian inference for source estimation}
\label{sec:Bayesian}
Following Bayes' rule, and the notation introduced by Arumpalam et al. for recursive Bayesian inference \cite{arulampalam_tutorial_2002}, the posterior probability distribution of the emission rate $Q$ based on the measurements of $c^y$ at time step $k$ (or after the $k$'th sensor pass) is \cite{albertson_mobile_2016}, \cite{yee_bayesian_2007}
\begin{equation}
\label{eq:Bayes}
    p\left(Q_k\big|c^y_{1:k}\right) = \frac{p\left(Q_k \big|c^y_{1:k-1}\right)p\left(c^y_k\big|Q_k\right)}{p\left(c^y_{1:k}\right)},
\end{equation}
where $p\left(Q_k\big|c^y_{1:k}\right)$, $p\left(Q_k\big|c^y_{1:k-1}\right)$, and $p\left(c^y_{1:k}\right)$ are probability density functions (PDFs). $p\left(Q_k\big|c^y_{1:k-1}\right)$ is the prior that is being updated through the recursion, $p\left(c^y_k\big|Q_k\right)$ is the likelihood function, and $p\left(c^y_{1:k}\right)$ is the evidence term that ensures $p\left(Q_k\big|c^y_{1:k}\right)$ integrates to unity. Note that the notation $c^y_{a:b}$ refers to the contiguous set of measurements between time (sensor pass) $a$ and $b$ inclusive.

In practical applications, past measurements of similar facilities \cite{brantley_assessment_2014} may be used to formulate the prior probability distribution at the first time step, i.e. before any sampling activities. Under the assumption that the only prior knowledge of $Q$ is its lower and upper bounds, a uniform prior distribution can be adopted \cite{yee_bayesian_2007,yee_theory_2008}. This uniform distribution is often considered as sufficiently uninformative based on the principle of maximum entropy \cite{jaynes_prior_1968} and can be expressed as follows
\begin{equation}
\label{eq:prior}
    p(Q_1) = \frac{1}{Q_{max}-Q_{min}},
\end{equation}
where $Q_{max}$ and $Q_{min}$ are the prescribed upper and lower bounds of the emission rate, respectively. 

The likelihood function, $p\left(c^y_k\big|Q_k\right)$, describes the probability of observing $c^y$ given $Q$ at the $k$'th sensor pass and encodes all the information provided by the mass concentration measurements about the unknown emission rate \cite{yee_bayesian_2007}. Since the underlying distribution of the concentration measurements are unknown, the principle of maximum entropy supports the application of a Gaussian distribution with a prescribed error scale \cite{jaynes_probability_2003}. This choice for the likelihood function has proven useful in previous studies \cite{keats_bayesian_2007, yee_theory_2008, yee_inference_2010,albertson_mobile_2016}, and is therefore adopted here. Furthermore, the dataset investigated in this study has been previously tested for leak estimation using the Gaussian likelihood function with satisfactory results \cite{zhou_mobile_2019} (more details on the dataset are provided in section \ref{sec:Methods}) . Consequently,  The Gaussian likelihood function in this study is expressed as 
\begin{equation}
\label{eq:LF}
    p\left(c^y_k\big|Q_k\right) = \frac{1}{\sigma_e\sqrt{2\pi}}\exp\left[-\frac{1}{2}\left(\frac{c^y_k-c^{y,M}_k(Q_k)}{\sigma_e}\right)^2\right],
\end{equation}
where $c^{y,M}(Q) = \frac{Q}{u_e^M}D_z^M$ is the cross-plume integral of a modeled concentration for a given candidate value of $Q$. $D_z^M$ is the estimated value of $D_{z,e}$ based on a Lagrangian Stochastic Model (LSM). The LSM is used to describe plume dispersion in a turbulent flow by modeling paths of fluid particles, which are driven by the random velocity field modeled by the generalized Langevin equation \cite{wilson_review_1996}. In this study, we impose the so-called well-mixed conditions and adopt Thomson's simplest solution for statistically stationary and horizontally homogeneous turbulence \cite{thomson_criteria_1987}. The LSM takes meteorological measurements (friction velocity, surface roughness, standard deviation of $u$ and $w$ and Obukhov Length) and the estimated distance between the emission source and the sensor as input parameters. The LSM is previously described in further detail in \cite{zhou_mobile_2019}. 

In equation (\ref{eq:LF}), $\sigma_e$ is the uncertainty scale parameter, which can be estimated for observed data as
\begin{equation}
    \sigma_e = \sqrt{\frac{1}{N-1}\sum_{k=1}^{N} \left( c^y_k - c^{y,M}(Q_0) \right)^{2} },
\end{equation}
where $N$ is the number of passes per experiment. In this study, $\sigma_e$ is estimated from the controlled release experiments and is known prior to the application of the Bayesian inference approach. In cases where $\sigma_e$ cannot be estimated from prior measurements, it can be estimated using the error propagation method \cite{rao_uncertainty_2005,zhou_estimation_2019}. Briefly, $\sigma_e$ is due to the following error scale parameters: 1) error due to the stochastic nature of atmospheric plume dispersion, 2) error due to the plume dispersion model and 3) measurements error including errors from the model input data. The parameterization of each of these error parameters is dependent upon the local meteorological conditions, the dispersion model used, and the quality of measurements. 

The recursive Bayesian formulation of equation (\ref{eq:Bayes}) used in conjunction with the uniform prior of equation does not have an analytical solution and should be solved numerically. For the numerical solution, $Q$ is discretized from $Q_{min}$ to $Q_{max}$ with a uniform step size of $\Delta Q$ to form a vector of candidate values for $Q$ to be considered. For each measurement of $c^y$, the likelihood function is evaluated at all candidate $Q$ values using equation (\ref{eq:LF}) and multiplied by the prior probability distribution. Subsequently, the evidence term after the $k$'th sensor pass can be calculated through numerical integration as follows:
\begin{equation}
    p\left(c^y_{1:k}\right) = \sum_{Q_{min}}^{Q_{max}}  p\left(Q_k\big|c^y_{1:k-1}\right)p\left(c^y_k\big|Q_k\right) \Delta Q.
\end{equation}
After calculating the evidence term, equation (\ref{eq:Bayes}) can be applied to evaluate the posterior distribution of the emission rate at each candidate $Q$ value. The same procedure is repeated after each mobile sensor pass and the posterior distribution is updated using a new prior (posterior at previous sensor pass) and a newly calculated likelihood function with the most recent $c^y$ measurement. 

After each sensor pass, the posterior PDF can be used to estimate the emission rate and the associated uncertainty. For instance, the emission rate after sensor pass $k$, can be calculated as the mean, median or mode of of the posterior PDF $p\left(Q_k\big|c^y_{1:k}\right)$. Given that an uninformative, uniform prior distribution was adopted, we expect that the median and the mean to be heavily affected by the prior in the early stages of analysis. Therefore, the mode of the posterior PDF $p\left(Q_k\big|c^y_{1:k}\right)$ is used as the estimated emission rate, i.e.:
\begin{equation}
    E_k^Q = \argmax_{Q_k}  \left[ p\left(Q_k\big|c^y_{1:k}\right) \right].
\end{equation}

As number of mobile passes are increased, the effects of the prior distribution are reduced and the mode, median and mean of the posterior PDF grow closer in value. Furthermore, in cases where an informative prior can be derived prior from past experiments, the mean or median of the posterior PDF may be better candidates for the emission rate, as they better incorporate the prior information than the mode. 

Finally, the associated uncertainty of the emission rate estimation using Bayesian inference is often calculated as the standard deviation, $\sigma_k^Q$, of the posterior PDF:
\begin{equation}
    \left(\sigma_k^Q\right)^2 = \int_{Q_{min}}^{Q_{max}} \left(Q-\Bar{Q_k}\right)^2 \times p\left(Q_k\big|c^y_{1:k}\right) dQ,
\end{equation}
where $Q_k$ is the expectation of the posterior PDF evaluated as follows:
\begin{equation}
    \Bar{Q_k} = \int_{Q_{min}}^{Q_{max}} Q \times p\left(Q_k\big|c^y_{1:k}\right) dQ.
\end{equation}

\subsection{Bayesian inference for changepoint detection}
\label{sec:Fault}
In order to detect a change in the source emission rate, we apply the Bayesian Online Changepoint Detection (BOCD) methodology \cite{adams_bayesian_2007}. In this approach, changepoints are found by first estimating the posterior distribution over the run length, i.e. the time (or in this case, number of sensor passes) since the last changepoint, given the data observed so far. Denoting the length of the current run after sensor pass $k$ with $r_k$, and applying the definition of conditional probability the run length posterior distribution can be expressed as
\begin{equation}
    p\left(r_k\big|c^y_{1:k}\right) = \frac{p\left(r_k, c^y_{1:k}\right)}{p(c^y_{1:k})},
\end{equation}
where the run length evidence term is calculated using $p(c^y_{1:k}) = \sum_{r_k} p\left(r_k, c^y_{1:k}\right)$.

After every sensor pass there are two possibilities regarding the changepoint: 1) No changes occur after the sensor pass and therefore the run length is increase by 1 or 2) change occurs and run length is reset to 0. The probability that no changepoint occurs is referred to as the "growth probability" as it indicates that the run length is growing by 1 compared to the previous sensor pass. Similarly, we refer to the probability that a changepoint occurs as the "changepoint probability". We denote the growth probability such that the run length reaches $i \in \{0,1,2, ..., k-1\} $ after $k$ sensor passes by $\alpha_k(i)$ and derive a recursive estimate as follows

\begin{align}
\label{eq:gp1}
    \alpha_k(i) &= p\left(r_k=i, c^y_{1:k}\right) \nonumber \\
    &= p\left(r_k = i, r_{k-1} = i-1, c^y_k, c^y_{k-1}\right) \nonumber \\
    &= p\left(r_k = i, c^y_k\big|r_{k-1} = i-1, c^y_{1:k-1}\right) p\left(r_{k-1}=i-1, c^y_{1:k-1}\right) \nonumber \\
    &= p\left(r_k = i, c^y_k\big|r_{k-1} = i-1, c^y_{1:k-1}\right) \alpha_{k-1}(i-1).
\end{align}

The first step in equation (\ref{eq:gp1}) is due to the fact that the run length can only increase by 1 after each sensor pass and the second step follows from the chain rule in probability theory. Furthermore, the recursive estimate in equation (\ref{eq:gp1}) requires the evaluation of $p\left(r_k = i, c^y_k\big|r_{k-1} = i-1, c^y_{1:k-1}\right)$ which can be achieved by employing the chain rule again:

\begin{align}
\label{eq:gp2}
    p\left(r_k, c^y_k\big|r_{k-1}, c^y_{1:k-1}\right) &= p\left(r_k\big|r_{k-1}, c^y_{1:k-1}, c^y_{k}\right) p\left(c^y_k\big|r_{k-1}, c^y_{1:k-1}\right) \nonumber \\
    &= p\left(r_k|r_{k-1}\right) p\left(c^y_k\big|c^{y}_{k-i:k-1}\right),
\end{align}
where $r_k = i$ and $r_{k-1}=i-1$ are implied and not explicitly written for simplicity of notation. The final step of equation (\ref{eq:gp2}) follows from the independence of the measured mass concentrations before and after a changepoint, which is true based on our assumption of independence of leak rates before and after a changepoint. Further, the condition on $r_k$ in the second term on right hand side of the equation is absorbed by only limiting the conditional probability on measurements since the last changepoint leading to the subscript $k-i:k-1$. Equation (\ref{eq:gp2}) suggests that the growth probability can be computed based on two calculations: 1) The prior over $r_k$ given $r_{k-1}$ (also referred to as the changepoint prior) and 2) The predictive distribution over the new measurement after sensor pass $k$, given the data since the last changepoint. 

The changepoint prior in equation (\ref{eq:gp2}) has nonzero mass at only two outcomes, because after each sensor pass the run length either continues to grow such that $r_k = r_{k-1}+1$ or a changepoint occurs and $r_k = 0$. Furthermore, the hazard function can be used to quantify each of these outcomes, since by definition the hazard function quantifies the probability that a changepoint occurs at a given time step conditioned that no changepoint has occurred prior to that time step \cite{adams_bayesian_2007,forbes_statistical_2010}. Therefore, the changepoint prior can be expressed as
\begin{equation}
    p\left(r_k|r_{k-1}\right) = 
    \begin{cases} 
      H\left(r_{k-1}+1\right) & \text{if}\; r_k = 0 \\
      1-H\left(r_{k-1}+1\right) & \text{if}\;r_k = r_{k-1}+1 \\
      0 & \text{otherwise},
   \end{cases}
\end{equation} 
where $H\left(r_{k-1}\right)$ is the hazard function. The hazard function depends on the discrete a priori probability distribution over the interval between changepoints. However, in this study, we consider the special case where the a priori probability distribution is a discrete geometric distribution with timescale $\lambda$, i.e. the run length distribution is due to a memoryless process and the hazard function is constant at $1/\lambda$. The timescale $\lambda$ can be set through prior knowledge, for instance, the average number of passes completed before a change in the emission rate occurs. In this study, $\lambda$ is set to 15 based on the conducted experiments described in section \ref{sec:Experiments}.

The predictive distribution $p\left(c^y_k\big|c^{y}_{k-i:k-1}\right)$ can be described by an equivalent distribution by utilizing the plume transport model of equation (\ref{eq:cy2}) leading a one-to-one correspondence between the predictive distribution $p\left(c^y_k\big|c^{y}_{k-i:k-1}\right)$ and $p\left(Q_k\big|c^{y}_{k-i:k-1}\right)$. In this case, the predictive distribution $p\left(c^y_k\big|c^{y}_{k-i:k-1}\right)$ can be found through scaling of $p\left(Q_k\big|c^{y}_{k-i:k-1}\right)$. We note that $p\left(Q_k\big|c^{y}_{k-i:k-1}\right)$ is the prior distribution in equation (\ref{eq:Bayes}), given the data since the last changepoint. Therefore, $p\left(Q_k\big|c^{y}_{k-i:k-1}\right)$ and consequently $p\left(c^y_k\big|c^{y}_{k-i:k-1}\right)$ can be estimated using the recursive Bayesian approach detailed in section \ref{sec:Bayesian}, leading to simultaneous estimation of leak rate and changepoint detection. The computational details of calculating the growth probability term are provided in section \ref{sec:Comp}.

The changepoint probability is evaluated in a similar manner to the growth probability. By noting that with the occurrence of a changepoint, the run length $r_k$ drops to 0, we can derive a recursive estimate for the changepoint probability as

\begin{align}
\label{eq:cp}
    \alpha_k(0) &= p\left(r_k=0, c^y_{1:k}\right) \nonumber \\
    &= \sum_{j=1}^{k-2} p\left(r_k=0, r_{k-1} = j, c^y_k, c^y_{k-1}\right) \nonumber \\
    &= \sum_{j=1}^{k-2} p\left(r_k = 0 | r_{k-1} = j\right) p\left(c^y_k\big|c^y_{k-j-1:k-1}\right)\alpha_{k-1}(j) 
\end{align}
where $\alpha_k(0)$ is the changepoint probability. The summation after the first step of equation (\ref{eq:cp}) appears due to marginalization over the run length at sensor pass $k-1$. The intermediate steps in the derivation are omitted as they are identical to the derivation of the growth probability as outlined in equations (\ref{eq:gp1}) and (\ref{eq:gp2}). The estimation of each term on the right hand side of equation (\ref{eq:cp}) follows the same procedure as equation (\ref{eq:gp2}).

\begin{figure}
\centering
  \includegraphics[scale=0.65]{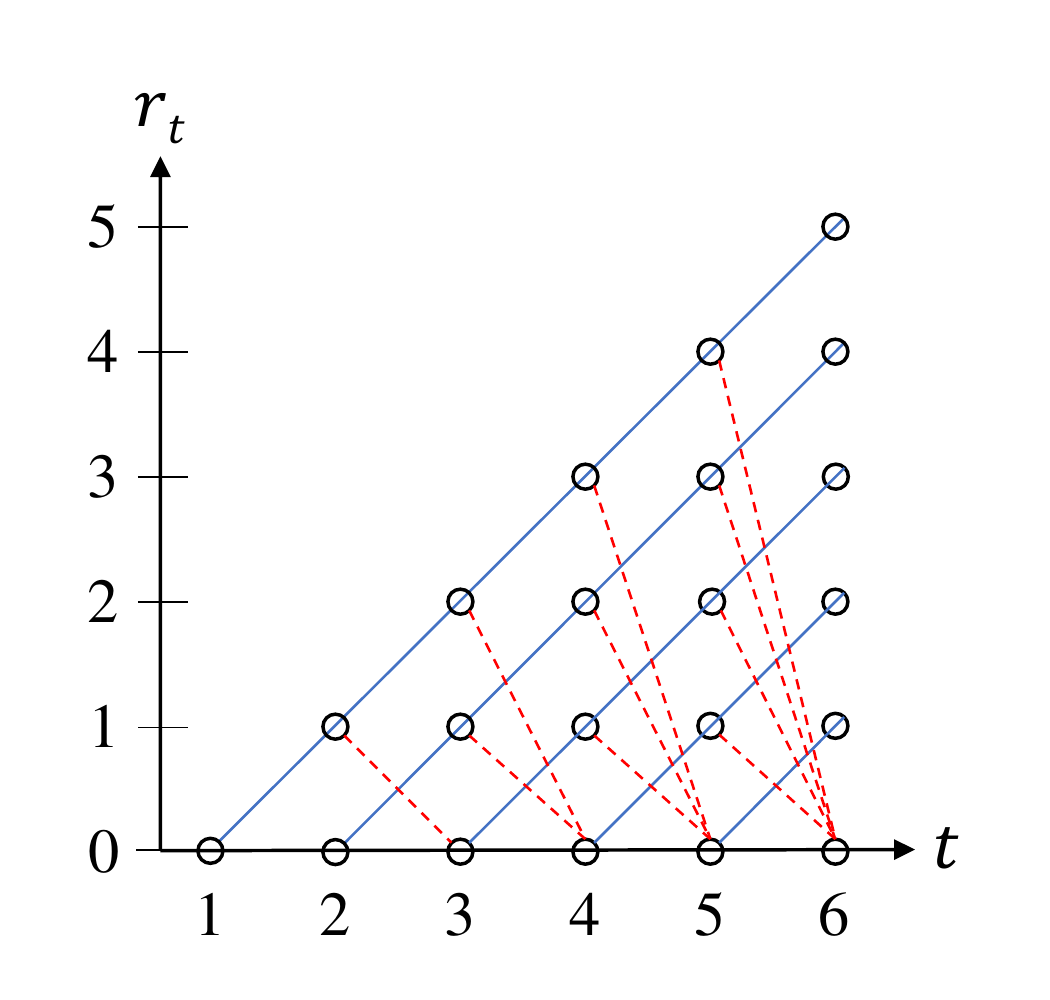}
  \caption{Visual description of the message passing algorithm used to estimate the run length distribution over the observed data. The circles represent run length hypotheses and the lines between the circles show recursive transfer of mass between sensor pass (or time step). Solid lines indicate that probability mass is being passed upwards, causing the run length to grow at after the next sensor pass and dashed lines indicate that the current run is truncated, and the run length drops to zero.}
  \label{fgr_trellis}
\end{figure}

Figure \ref{fgr_trellis} illustrates the algorithm used to estimate the posterior distribution over the run length as described by equations (\ref{eq:gp2}) and (\ref{eq:gp1}). In this diagram, the solid blue lines correspond to growth probability calculations, while the red dashed lines are associated with changepoint probability evaluations. We note that multiple dashed lines arriving at a node in Figure \ref{fgr_trellis}, correspond to the marginalization over the run length at the previous sensor pass. 

In practice, to automatically detect a changepoint and alert the system that a change in leak rate has occurred a prescribed condition on the changepoint probability should be put into place. The appropriate detection condition can be chosen based on the application and the available ancillary information regarding the emission conditions. In this study, we use a changepoint probability threshold such that when the changepoint probability is above this threshold the system automatically registers a change in leak rate, the posterior probability over the emission rate at the previous sensor pass is retained and the prior distribution in equation (\ref{eq:Bayes}) is reset to the uniform prior of equation (\ref{eq:prior}).

\subsection{Computational details of growth probability estimation}
\label{sec:Comp}
The calculation of the growth probability distribution after sensor pass $k$ relies on knowledge of the probability distribution of the emission rate conditioned on the observed data from the last changepoint until the previous sensor pass at $k-1$, i.e., $p\left(Q_k\big|c^y_{k-i:k-1}\right)$ as shown by the derivation in equation (\ref{eq:gp2}). Furthermore, the growth probability is calculated for every value of run length $i \in \{0, 1, 2, ..., k-1\}$, which requires the recursive calculation of the posterior probability of emission rate for every $i$ through equation (\ref{eq:Bayes}). Therefore, the growth probability calculation can be computationally expensive for large values of $k$. To overcome this challenge, we note that the likelihood function in equation (\ref{eq:Bayes}) is independent of $i$, hence, the calculation of the posterior probability of emission rate for every $i$ should completed in a vectorized or Single Instruction, Multiple Data (SIMD) manner that allows for parallel processing of data and significantly improves the run time of the probability estimation \cite{flynn_computer_1972}.  

\section{Materials and Methods}
\label{sec:Methods}
In this section we first describe the conducted field experiments  that yielded the data used to examine the Bayesian framework. Throughout these experiments, the leak rate was kept constant, therefore, a data synthesis procedure is implemented to simulate a step change in the leak rate before the application of the changepoint detection algorithm. A series of performance measures are then defined to quantify the performance of the changepoint detection algorithm.

\subsection{Field experiments}
\label{sec:Experiments}
Experiments of controlled releases of methane were conducted at the McGovern soccer training field of Cornell University (Game Farm Rd, Ithaca, NY. USA) in early August 2016. During the experiments, the site was covered with short grass ($\sim$5cm) and located in a relatively open field with a ~500m distance from a residential area in the West, ~ 150m distance from small forest in the North and approximately 400m (500m) distance from roads on the east (south) side. Point-source emission of methane (99.9\% pure gas) was controlled by a mass flow controller (SmarTrak 100 from Sierra Instruments Inc., Monterey, CA, USA), with a mass flow accuracy of $\pm 1 \%$. The height of the methane release was similar to the height of the grass at ~5cm. On the west side of the field, a 3D sonic anemometer (CSAT-3, Campbell Scientific Inc., Logan, UT, USA) was installed on a small tower to measure local meteorological conditions. The height of the tower was 2.31m with the sonic anemometer measuring the three components of wind velocity and air temperature at a frequency of 10 Hz. To mimic emissions from a source surrounded by other low-level structures, a 1.4m barrier (windbreak) was established in a circle around the emission source. This setup can for instance, approximate a well head located in densely organized well pad or a pipeline within a natural gas metering station. 

A mobile measurement platform (MMP) was configured with a precise GPS unit (Trimble Geo 7X handheld from Trimble Inc., Sunnyvale, CA, USA) to track its position at a sampling frequency of
1 Hz. The accuracy of the GPS unit was approximately 5-15 cm for $>$97\% of the measured data points after post processing. The MMP was equipped with a LI-COR LI-7700 open-path methane analyzer (LI-COR Biosciences, Lincoln, NE, USA), that outputs methane mixing ratios in the unit of parts per million (ppm). The operating frequency of the analyzer was set to 10 Hz, and it was positioned at a height of 1.3m. Furthermore, the analyzer was calibrated by the manufacturer less than a month before the experiment, and is designed with an open-path configuration for long term monitoring without regular re-calibration.

Regarding the measurements, a conversion factor is applied to translate above-ambient mixing ratios ($c_a$, in ppm) into mass concentrations ($c$, in g/m\textsuperscript{3}) in equation (\ref{eq:F}). This conversion factor is dependent on the molecular weight of the released gas (16.04 g/mol for methane) and the ambient temperature which affects the molar of the gas. The above-ambient mixing ratios are found by subtracting the ambient methane mixing ratios from the raw methane mixing ratios measured by the open-path analyzer, with the ambient mixing ratio calculated as the 5th percentile of the ranked time series of raw mixing ratio measurements \cite{brantley_assessment_2014,foster-wittig_estimation_2015,zhou_mobile_2019,albertson_mobile_2016}. The estimated ambient mixing ratio was compared to methane mixing ratios measured prior to the experiments with minimal differences found ($<$2\%), suggesting that the ambient mixing ratio was determined robustly.  

To ensure perpendicular sensor passes with respect to the wind direction, stake flags were placed in three circles centered at the emission source with radii of 10, 20 and 30m, and repeated passes were made along each of the circles. The average sensor speed was very low during the experiments (approximately 2 m/s) to better capture the plume structure. In addition, the start of a pass took place approximately one minute after the end of the previous pass, warranting the independence of the measurements of each pass from those of previous passes. Data was aggregated within 30 minute intervals, during which the measurements from the meteorological tower was used to estimate meteorological parameters such as the Obukhov length ($L$) and the friction velocity ($u_*$). Each set of passes completed in a 30-minute period is considered and analyzed as a single experiment. 
\subsection{Data synthesis}
\label{sec:Data}
A total of 18 experiments were conducted with the 1.4m barrier present in the field, with six experiments at each source-to-sensor distance (i.e. $x_m$ of 10, 20 and 30m). Two selection requirements were established to filter out experiments performed under unacceptable conditions. First, experiments under stable atmospheric conditions were excluded, and only experiments conducted under neutral or unstable conditions were retained. Second, experiments conducted under low wind ($\Bar{u} < $ 1.0 m/s) and high turbulent intensity ($I_u>$0.5) conditions were discarded. As a result of this selection criteria,  14 experiments were retained for further analysis, the details of which are summarized in Table \ref{Table1} alongside the meteorological conditions reported by the meteorological tower.

\begin{table}
    \caption{Summary of experimental conditions, including experiment identification number (ID), approximate source-to-sensor distance ($x_m$), number of sensor passes for the experiment ($N$), and sampling day of year (DOY), and meteorological conditions as measured by the nearby meteorological tower, including mean streamwise velocity ($\Bar{u}$), standard deviation of streamwise velocity ($\sigma_u$), turbulent intensity ($I_u$), friction velocity ($u_*$), mean wind direction ($\theta_m$) clockwise from north, sensible heat flux ($H$), and atmospheric stability ($z/L$). The meteorological variables are derived from data collected during the corresponding experiments (~30 min).}
    \label{Table1}
\centering
\begin{tabular}{l|cccccccccc}
\hline
\textbf{ID} & \textbf{$x_m$ (m)} & $N$ & \textbf{DOY} & \textbf{$\Bar{u}$ (m/s)} & \textbf{$\sigma_u$ (m/s)} & \textbf{$I_u$ (-)} & \textbf{$u_*$ (m/s)} & \textbf{$\theta_m$ (deg)} & \textbf{$H$ (W/m\textsuperscript{2})} & \textbf{$z/L$ (-)} \\
\hline
1  & 30        & 14  & 217 & 2.94            & 0.98             & 0.28      & 0.18        & 147              & 77.67                        & -0.32      \\
2  & 20        & 16  & 217 & 2.72            & 1.06             & 0.31      & 0.23        & 149              & 148.97                       & -0.31     \\
3  & 10        & 15  & 217 & 2.49            & 0.98             & 0.33      & 0.29        & 172              & 138.13                       & -0.14     \\
4  & 30        & 12  & 217 & 2.72            & 1.15             & 0.31      & 0.24        & 152              & 161.21                       & -0.30     \\
5  & 20        & 16  & 217 & 2.95            & 1.08             & 0.29      & 0.22        & 146              & 171.01                       & -0.41     \\
6  & 10        & 16  & 217 & 2.48            & 0.97             & 0.33      & 0.21        & 159              & 159.46                       & -0.46     \\
7  & 30        & 14  & 218 & 3.41            & 1.34             & 0.28      & 0.37        & 204              & 219.77                       & -0.11     \\
8  & 20        & 14  & 218 & 3.70            & 1.27             & 0.26      & 0.36        & 211              & 223.92                       & -0.12     \\
9  & 10        & 13  & 218 & 3.82            & 1.33             & 0.26      & 0.38        & 212              & 207.86                       & -0.10     \\
10 & 30        & 16  & 218 & 4.18            & 1.31             & 0.24      & 0.36        & 194              & 171.36                       & -0.09     \\
11 & 20        & 13  & 218 & 4.31            & 1.31             & 0.24      & 0.37        & 184              & 173.08                       & -0.08     \\
12 & 10        & 13  & 218 & 3.99            & 1.23             & 0.25      & 0.40        & 179              & 129.53                       & -0.05     \\
13 & 20        & 12  & 219 & 2.75            & 1.04             & 0.30      & 0.16        & 318              & 125.83                       & -0.77     \\
14 & 10        & 13  & 219 & 2.29            & 1.04             & 0.35      & 0.20        & 315              & 140.29                       & -0.47     \\
\hline
\end{tabular}
\end{table}

\begin{figure}
\centering
  \includegraphics[scale=0.67]{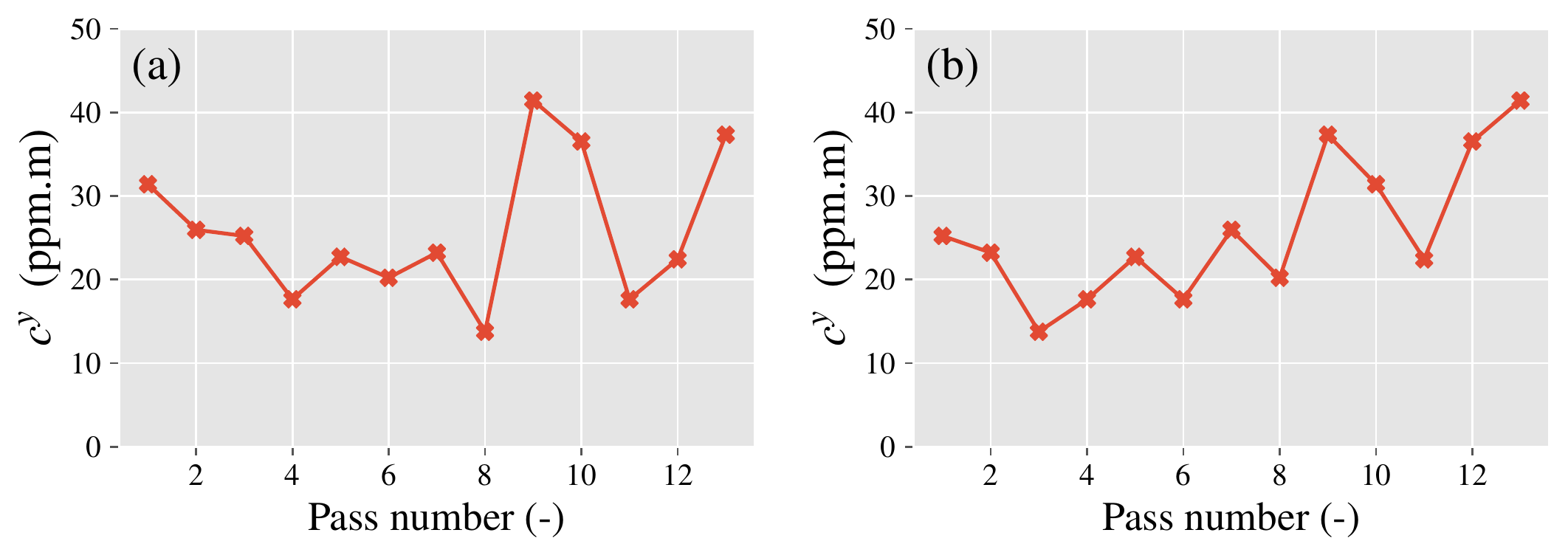}
  \caption{Two instances (a) and (b) of the same experiment (experiment ID 4) created through random shuffling of the measurements. The measurements are cross-plume integrated above-ambient mass concentrations of methane calculated over sensor passes.}
  \label{fgr_permutation}
\end{figure}

In all the conducted experiments, the emission rate was kept constant at a rate of $Q_0$ = 0.083 g/s. Therefore, a data synthesis procedure was established to artificially simulate a change in the emission rate. The proposed data synthesis approach relies on the observation that the order of the measurements of the cross-plume integrated mass concentrations through each pass of the MMP can affect the performance of the changepoint detection algorithm. For instance, consider two permutations of the measurements of the same experiment (experiment ID 14 in Table \ref{Table1}) as shown in Figure \ref{fgr_permutation}. In the first permutation,  series of measurements leading to multiple low $c^y$ values are followed by a comparatively high measurement (at sensor pass 9), therefore the algorithm is likely to detect a changepoint while in reality the emission rate has been constant. However, in the second permutation, the difference in consecutive $c^y$ measurements are smaller than the first permutation and the probability of detecting a changepoint is lowered.

\begin{figure}
\centering
  \includegraphics[scale=0.55]{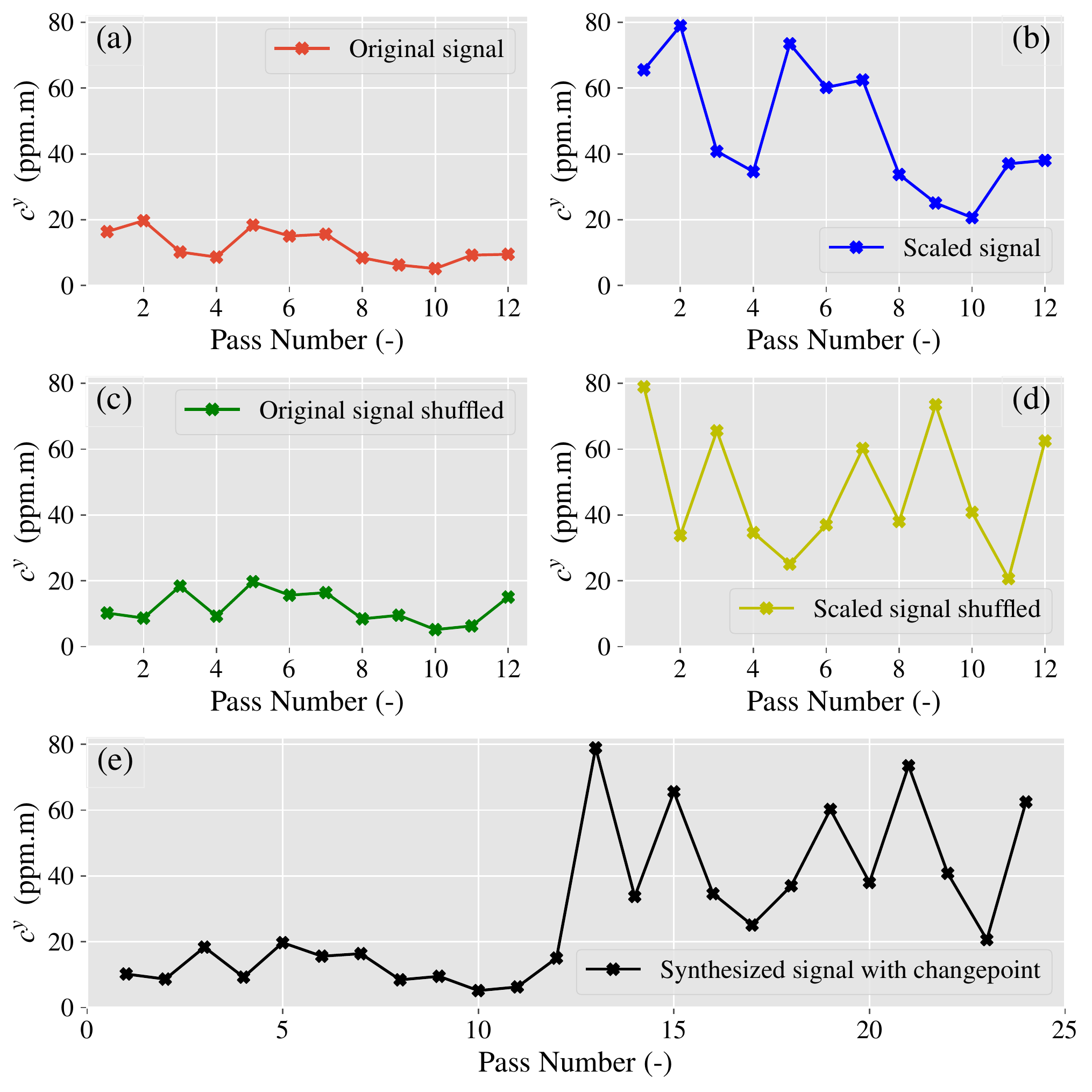}
  \caption{Summary of the data synthesis procedure for one instance of an experiment (ID 4), where the (a) original measurements from the field experiment are (b) scaled by multiplying all measurements by a prescribed constant. Two sets of new measurements are created by random shuffling leading to a (c) shuffling of the original measurements and a (d) shuffling of the scaled measurements. The two shuffled sets are then (e) concatenated to create one instance of synthesized measurements consisting of a step change.}
  \label{fgr_synth}
\end{figure}

This observation motivates the use of random shuffling in synthesizing experiments consisting of a change in the leak rate after a few passes of the MMP around the point source. For each experiment, the data synthesis steps are as follows (Experiment ID 4 in Table \ref{Table1} used in Figure \ref{fgr_synth}):
\begin{enumerate}
\item For each pass, $c^y$ is calculated to create a time series for the experiment, which will be referred to as the "original signal" as depicted in Figure \ref{fgr_synth}a.
\item The original signal is duplicated and scaled by a given constant that is chosen based on the ratio of the leak rate before and after a simulated changepoint resulting in a "scaled signal" as presented in Figure \ref{fgr_synth}b.
\item Both original and scaled signals are shuffled to create a random permutation of each respective signal with the results shown in Figure \ref{fgr_synth}c and \ref{fgr_synth}d
\item The two shuffled signals are concatenated such that the first measurement of the shuffled scaled signal follows the last measurement of the shuffled original signal, creating a signal that consists of a changepoint as illustrated in Figure \ref{fgr_synth}e.
\end{enumerate}

Steps 2-4 of the above procedure are repeated 1000 times, to create a total of 1000 signals with changepoints for each experiment. The changepoint detection algorithm is then applied to these signals and the performance of the algorithm is evaluated according to the performance measures described in section \ref{sec:Performance}. 

\subsection{Performance measures}
\label{sec:Performance}
The performance of changepoint detection methods are often evaluated through a series of commonly used measures. The importance of each measure is dependent upon the application of the changepoint detection system. Here, we introduce four different performance measures, that are slightly altered with respect to common definitions to better suit the context of changepoint detection in emission rates. In the following description of the performance measures, each unique signal consisting of a changepoint is referred to as an "instance" of an experiment.

\paragraph{Recall}
This refers to the portion of the changepoints that are detected after exactly one sensor pass following the change in emission rate. We label these successful changepoint detections as "True Positive" instances denoted by TP. Instances where the changepoints are detected with a delay, i.e. where changepoints are detected after at least two sensor passes after the change in emission rate are labeled "Delayed True Positive" (DTP), and instances where changepoints were not detected are referred to as "False Negative" (FN) instances. Therefore, recall which is a measure of how effective the changepoint algorithm is in detecting changepoints as soon as they occur is expressed as
\begin{equation}
    Recall = \frac{TP}{TP + DTP + FN}.
\end{equation}

\paragraph{Detection Recall}
This refers to the portion of changepoints that are detected any time after the change in emission rate has occurred. Employing the labels introduced earlier, detection recall as a measure of how effective the changepoint algorithm is in detecting the changepoints is expressed as
\begin{equation}
    Detection \; Recall = \frac{TP+DTP}{TP+DTP+FN}.
\end{equation}

\paragraph{Detection Delay}
This measures the average number of passes that it takes to detect the changepoint after the emission rate has changed. This measure is evaluated only for experiments where the changepoints were detected for all instances (with or without delay) and is evaluated as
\begin{equation}
    Detection \; Delay = \frac{\sum_{i=1}^{TP+DTP} Predicted \; CP - Actual \; CP}{TP + DTP},
\end{equation}
where $Predicted \; CP$ refers to the sensor pass after which the changepoint is detected and $Actual \; CP$ refers to the sensor pass after which the change in leak rate has occurred. 

\paragraph{False Positive Rate}
This refers to the ratio of number of instances where changepoints are detected prior to the change in emission rate to to total number of instances. Here "False Positive" (FP) refers to instances where data points that are not changepoints are recognized as changepoints. The False Positive Rate is a measure that reflects how many false alarms would be generated by the changepoint detection algorithm and is expressed as follows
\begin{equation}
    False\;Positive\;Rate = \frac{FP}{All\;instances} = \frac{FP}{1000},
\end{equation}
noting that $All \; instances$ refers to the 1000 signals with changepoints created for each experiment.  

To account for the variability introduced through random shuffling during the data synthesis stage, for each experiment the data synthesis procedure is repeated 100 times and therefore 100 different estimates of each performance measure are computed. These 100 values of the performance measures are then collected and used in a bootstrapping significance test analysis to establish 95\% confidence intervals for the computed performance measures \cite{efron_bootstrap_1986}. 

\section{Results and Discussion}
\label{sec:Results}
Before presenting the results related to the performance of the changepoint detection algorithms across all experiments, we explore one instance of an experiment. For this instance, it is shown how the changepoint detection algorithm is coupled with the emission estimation procedure to approximate the leak rates before and after a change in the emission rate. 
\subsection{Leak estimation and changepoint detection}
\label{sec:Res1}
\begin{figure}
\centering
  \includegraphics[scale=0.68]{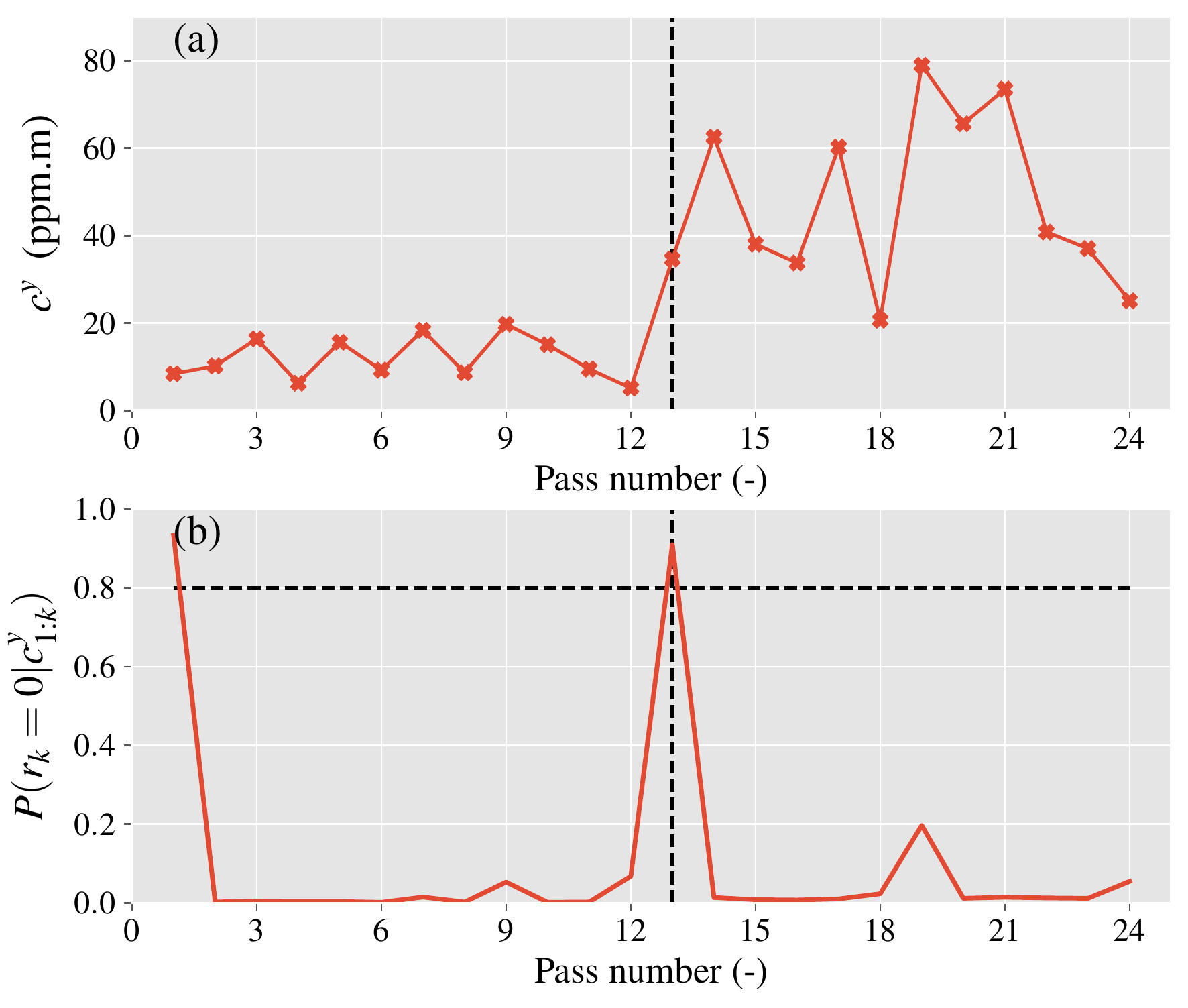}
  \caption{Application of the changepoint algorithm to an instance of an experiment (ID 4). (a) Synthesized instance with a step change in leak rate from 0.083 g/s to 0.332 g/s, where the vertical dashed line indicates the first sensor pass after the change. (b) The changepoint probability plotted after every sensor pass, where the horizontal dashed line represents the changepoint threshold probability, above which the algorithm registers a changepoint and resets the recursive Bayesian inference for leak estimation.}
  \label{fgr_ExpCP}
\end{figure}

We first show the changepoint detection procedure for one instance of an experiment (Experiment ID 4 in Table \ref{Table1}). In this instance, 12 sensor passes are made before the leak rate is significantly increased from $Q_1$ = 0.083 g/s to $Q_2$ = 0.332 g/s which is four times as large as $Q_1$. The $c^y$ measurements including the changepoint are shown in Figure \ref{fgr_ExpCP}a. After each sensor pass, the changepoint probability is evaluated through the procedure described in section \ref{sec:Fault} with the values presented in Figure \ref{fgr_ExpCP}b. A changepoint is detected when the changepoint probability surpasses a prescribed probability threshold of 0.8, after which the prior to the recursive Bayesian inference of equation (\ref{eq:Bayes}) is reset to the uniform prior (equation(\ref{eq:prior})), so that the new emission rate can be approximated. Furthermore, the figure shows an increase in the changepoint probability after the 19th sensor pass which can be attributed to the high value of the $c^y$ measurement (in comparison to previous measurements) corresponding to this sensor pass. In this study, the changepoint probability threshold is chosen through trial and error to lower the false alarm rate of the detection algorithm, and its effect on false positive rate is investigated in section \ref{sec:Res2}.

\begin{figure}
\centering
  \includegraphics[scale=0.57]{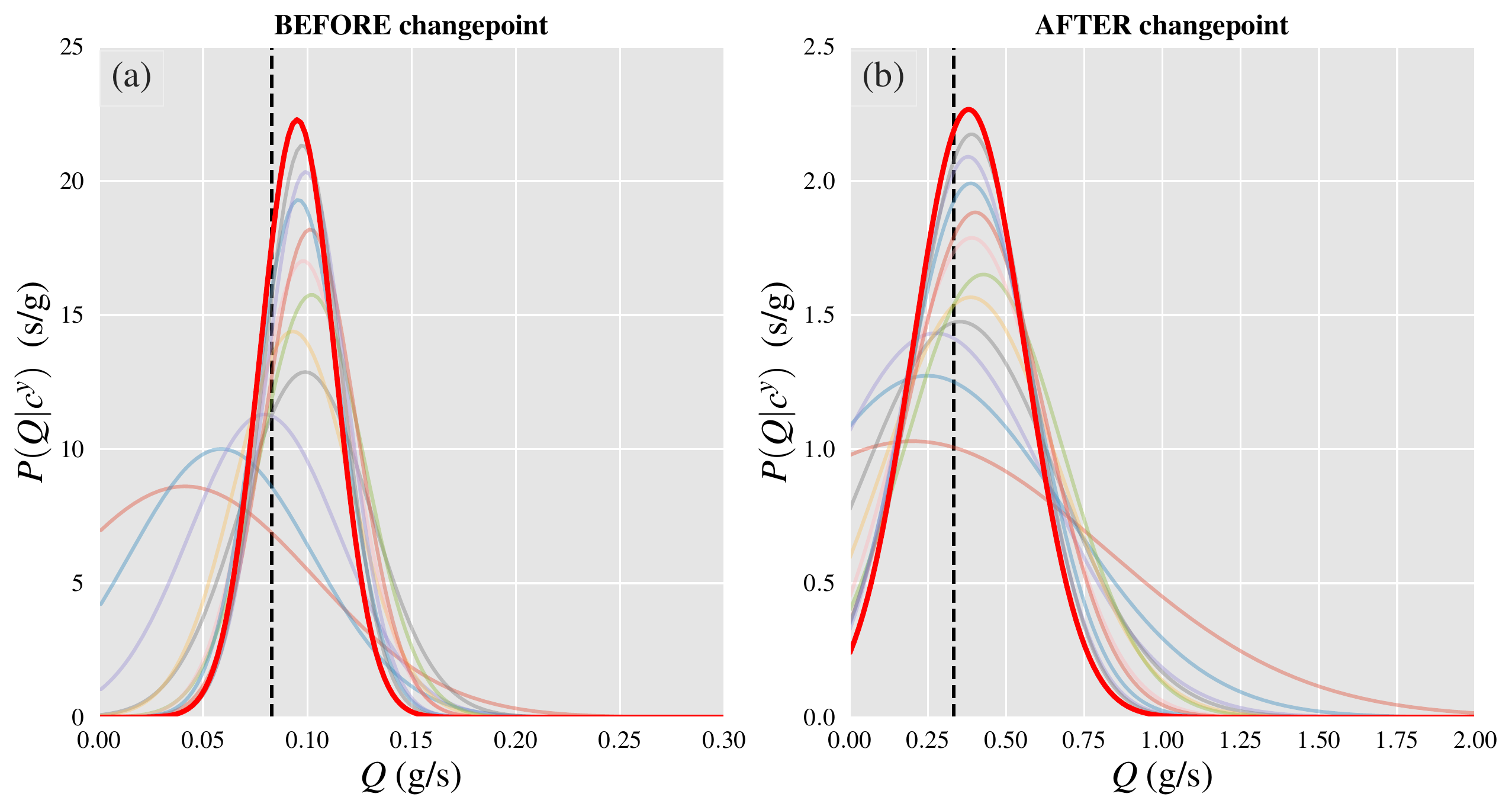}
  \caption{The evolution of the posterior probability $p\left(Q|c^y\right)$, of the emission rate $Q$ after each sensor pass (a) before and (b) after the change in leak rate as detected by the changepoint detection algorithm for one instance of an experiment (ID 4). The posterior probability obtained after the final sensor pass before the changepoint in (a) and after the overall final pass are presented with a solid red line. The vertical dashed lines indicate the actual emission rate of (a) 0.083 g/s and (b) 0.332 g/s.}
  \label{fgr_Post}
\end{figure}

For the same experiment instance as above, Figure \ref{fgr_Post} illustrates the evolution of the posterior PDF of the emission rate after each sensor pass before and after the changepoint. In this case, the lower and upper bounds of the emission rate, denoted by $Q_{min}$ and $Q_{max}$ are specified as 0 and 5.0 g/s. The choice for $Q_{min}$ is trivial as the emission rate can only take positive values. $Q_{max}$ is determined through trial and error such that the tail of the derived posterior PDF of the emission rate is close to zero. Using a larger $Q_{max}$ does not affect the accuracy of the Bayesian inference procedure, however it is deemed unnecessary as it increases the computational cost of the recursive Bayesian inference scheme. Figure \ref{fgr_Post}b shows that posterior PDF is fairly small at $Q$ = 2.0 g/s, suggesting that the choice of $Q_{max}$ = 5.0 g/s is effective. Starting from a relatively broad posterior PDF, suggesting a large uncertainty in the emission rate, the posterior PDF tends to approach a more narrow shape with additional sensor passes. It is worth noting that after the change in emission rate, the variation in the $c^y$ measurements are much larger compared to measurements at the original emission rate. Therefore, it is necessary to use a new estimate for $\sigma_e$ in equation (\ref{eq:LF}) for approximating the emission rate after the changepoint. In practice, the emission rate after the change is not known, hence, a larger and more conservative choice for $\sigma_e$ can be used to accommodate this lack of information. In the example shown in Figure \ref{fgr_Post}, we employ $\sigma_{e,2} = 10 \times \sigma_{e,1}$ where $\sigma_{e,1}$ and $\sigma_{e,2}$ refer to the error scale parameters before and after the changepoint, respectively, which is a conservative choice given that the error scale after the change is four times the error scale prior to the change. This conservative choice leads to a higher projected uncertainty when it comes to estimating the emission rate after the changepoint. 

\subsection{Changepoint detection performance}
\label{sec:Res2}
\begin{figure}
\centering
  \includegraphics[scale=0.56]{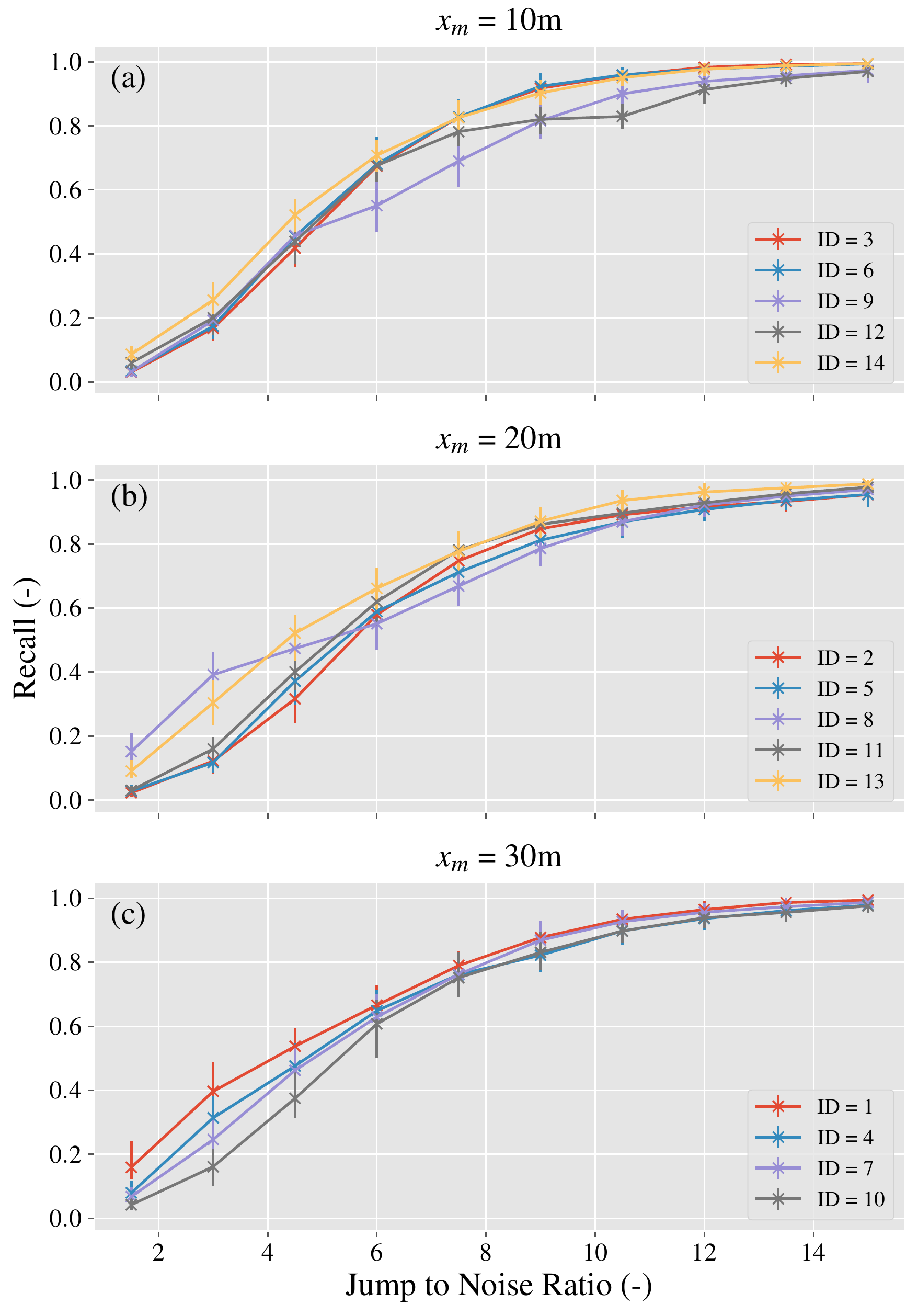}
  \caption{Evolution of recall for a series of jump to noise ratios varying between 1.5 and 15.5 for source-to-sensor distances, $x_m$ of (a) 10m, (b) 20m and (c) 30m. ID refers to the experiment ID as seen in Table \ref{Table1}. Vertical bars represent the 95\% confidence intervals.}
  \label{fgr_RecJNR}
\end{figure}

\begin{figure}
\centering
  \includegraphics[scale=0.60]{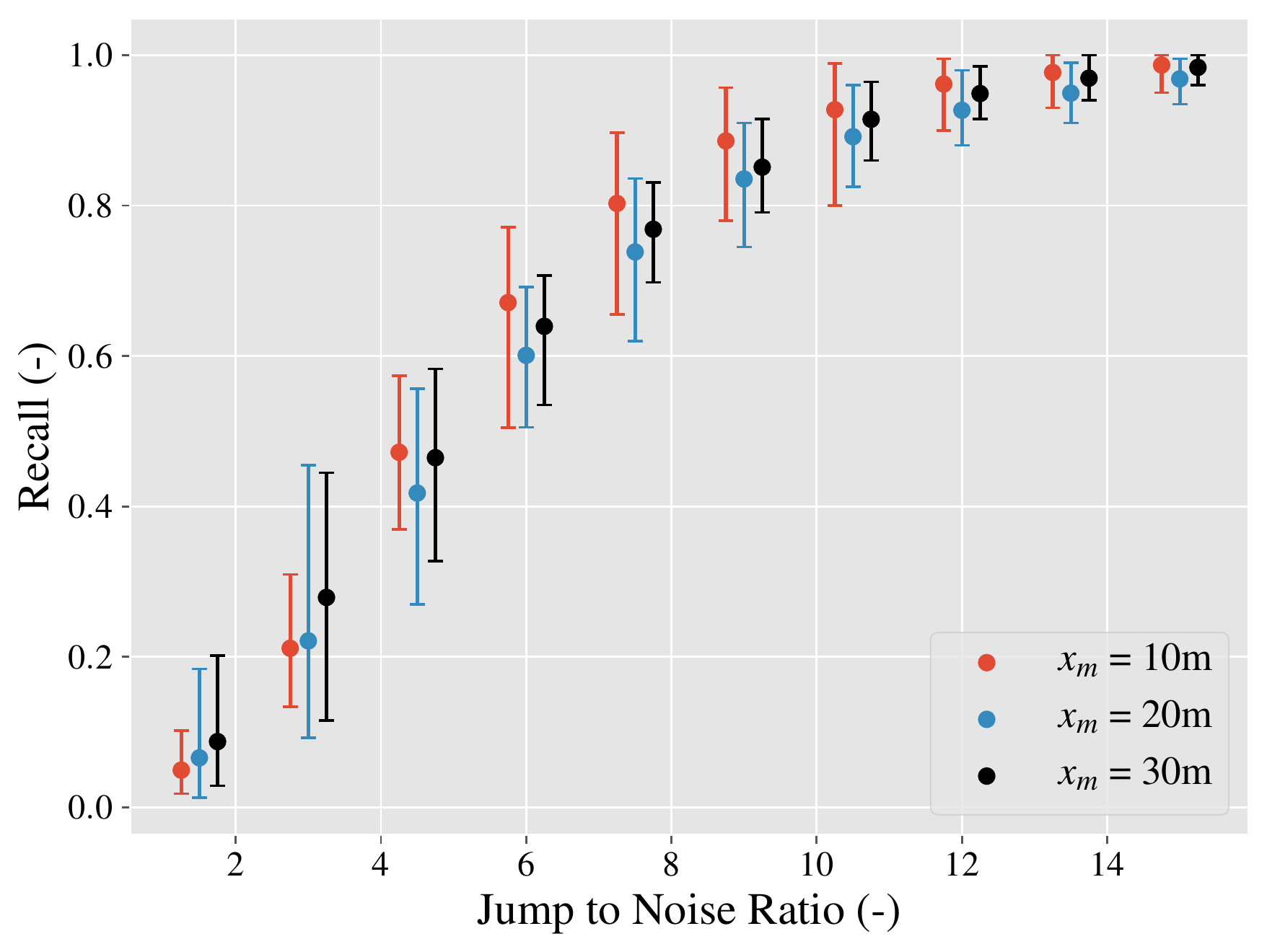}
  \caption{The evolution of recall for a series of jump to noise ratios varying between 1.5 and 15.5 after grouping experiments based on source-to-sensor distance, $x_m$. For each $x_m$, the set of jump to noise ratios are identical, however they are plotted in an offset to improve visibility. Vertical bars represent the 95\% confidence intervals. }
  \label{fgr_RecAll}
\end{figure}

We investigate the performance of the changepoint detection method using the measures introduced in section \ref{sec:Performance} by systematically varying the magnitude of the change in leak rate when synthesizing the data. To this end, Figure \ref{fgr_RecJNR} shows recall for varying values of "jump-to-noise ratio" (JNR), where JNR is the ratio of the absolute difference in the average $c^y$ before and after the change (i.e., the jump) to the standard deviation of $c^y$ before the change in leak rate (i.e., the noise). As expected, when the change in leak rate is of the order of the noise in the measurements, or in other words JNR is of the order of 1, changepoints are difficult to detect and therefore recall is low for all experiments. Further, as JNR is increased a monotonic rise in performance is observed across all experiments, with similar recall values observed in almost all cases. This similarity of recall values across experiments for each JNR motivates the idea of grouping all experiments based on the source-to-sensor distance. Figure \ref{fgr_RecAll} presents the recall averaged across all experiments within each group as a function of JNR, where the vertical bars indicate the 95\% confidence intervals. The trends observed in Figure \ref{fgr_RecJNR} and \ref{fgr_RecAll} suggest that JNR can solely predict the recall for the changepoint algorithm irrespective of the source-to-sensor distance and the measurement noise. 

\begin{figure}
\centering
  \includegraphics[scale=0.56]{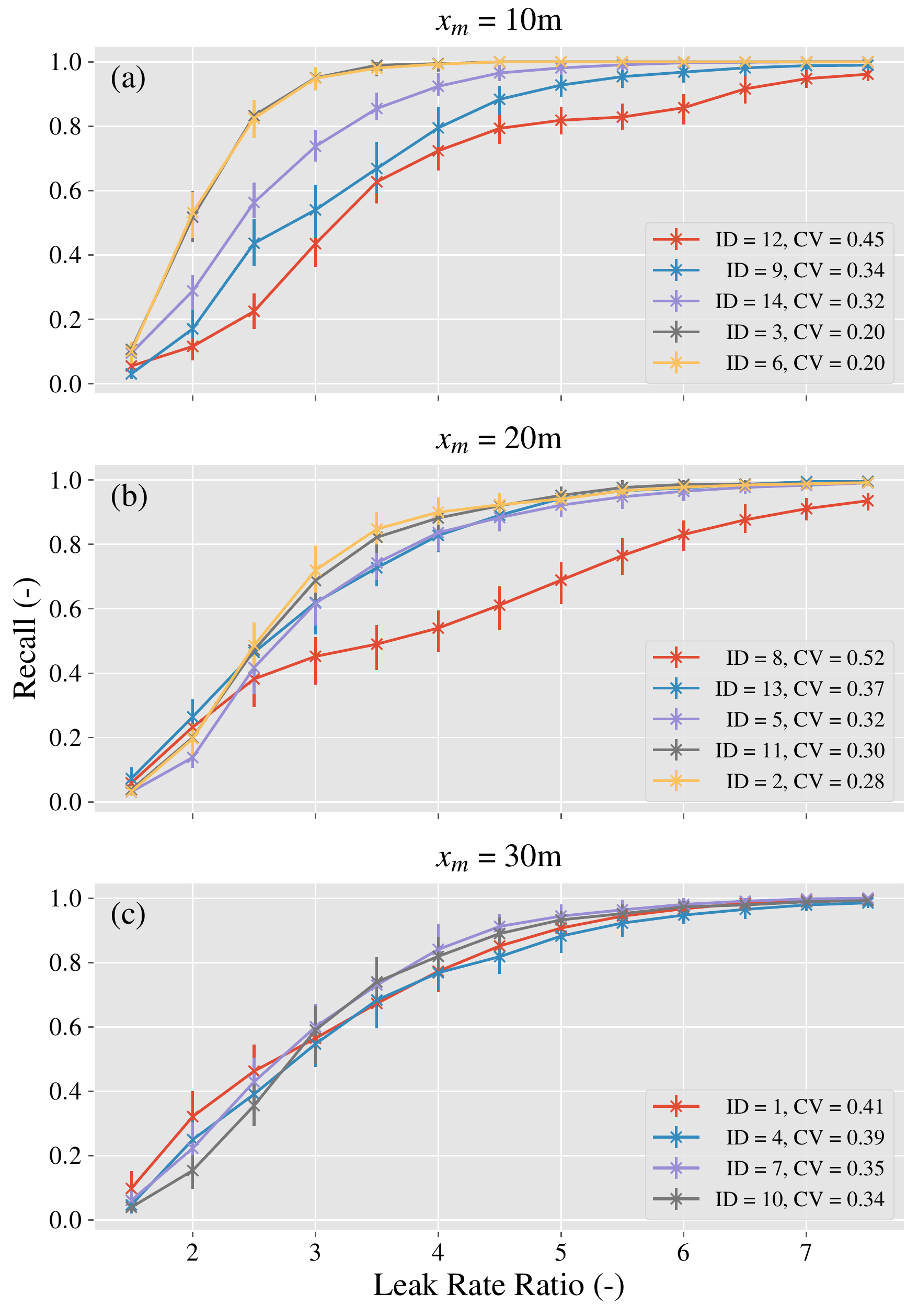}
  \caption{Evolution of recall for a series of leak rate ratios varying between 1.5 and 7.5 for source-to-sensor distances, $x_m$ of (a) 10m, (b) 20m and (c) 30m. ID refers to the experiment ID as seen in Table \ref{Table1}, and CV refers to coefficient of variation of $c^y$ measurements calculated for each experiment. Vertical bars represent the 95\% confidence intervals.}
  \label{fgr_RecQR}
\end{figure}

In practice, it is more constructive to predict the performance of the changepoint algorithm based on the ratio of the emission rate before and after the changepoint. To this end, Figure \ref{fgr_RecQR} illustrates recall as a function of increasing leak rate ratio (LRR) for all experiments, where leak rate ratio is the ratio of the leak rate after the change to the leak rate before the change. In this figure, for each source-to-sensor distance, the experiments are sorted based on coefficient of variation (CV) of $c^y$. For each experiment, CV is calculated as the ratio of the standard deviation of $c^y$ measurements to the average $c^y$ measurements in the original signal (e.g., Figure \ref{fgr_synth}a). It can be seen that a higher CV is a predictor for lower recall as an indicator for the performance of the changepoint algorithm. This relationship between recall and CV can be explained through a comparison between LRR and JNR.

According to equation (\ref{eq:cy2}), $c^y$ is directly proportional to the leak rate $Q$, therefore in our synthesized data, the leak rate ratio is the same as the ratio of the mean $c^y$ after and before the change. Therefore we can write
\begin{equation}
    LRR = \frac{Q_2}{Q_1} = \frac{\mu_{c_2}}{\mu_{c_1}},
\end{equation}
where $Q_2$ and $Q_1$ are the emission rate after and before the change, and $\mu_{c_2}$ and $\mu_{c_1}$ are the average $c^y$ measurements after and before the change, respectively. With this definition, we can relate JNR and LRR as follows
\begin{equation}
\label{eq:JNR}
    JNR = \frac{\mu_{c_2}-\mu_{c_1}}{\sigma_c} = \frac{LRR - 1}{CV},
\end{equation}
where $\sigma_c$ is the standard deviation of $c^y$ measurements in the original signal of an experiment. Based on equation (\ref{eq:JNR}), for a constant LRR, a higher value of CV corresponds to a smaller JNR, which according to Figure \ref{fgr_RecAll} points to a lower recall. 

\begin{figure}
\centering
  \includegraphics[scale=0.65]{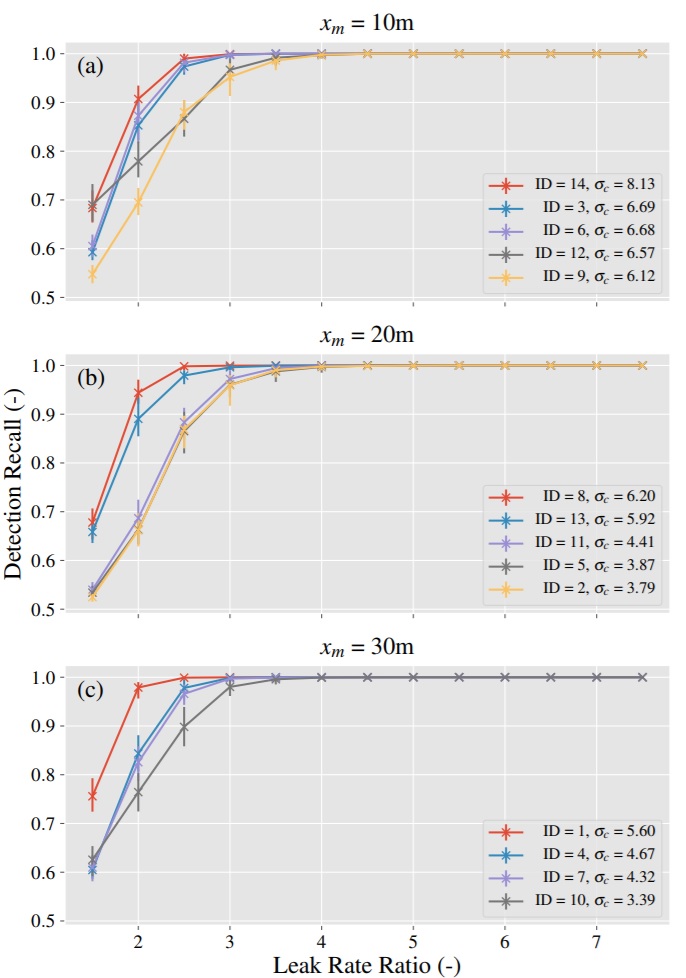}
  \caption{Evolution of detection recall for a series of leak rate ratios varying between 1.5 and 7.5 for source-to-sensor distances, $x_m$ of (a) 10m, (b) 20m and (c) 30m. ID refers to the experiment ID as seen in Table \ref{Table1}, and $\sigma_c$ refers to the standard deviation of $c^y$ measurements before the changepoint that is calculated for each experiment. Vertical bars represent the 95\% confidence intervals.}
  \label{fgr_DetRec}
\end{figure}

In most practical applications, delayed detection of the changepoint is acceptable. Therefore, Figure \ref{fgr_DetRec} depicts the detection recall as a function of increasing LRR. In this figure, for each source-to-sensor distance, the experiments are sorted based on $\sigma_c$ as a measure of noise in the measurements. It can be seen that experiments with higher values of $\sigma_c$ correspond to higher detection recalls. This behaviour is expected due to our data synthesis procedure, where high values of $\sigma_c$ lead to significantly large $c^y$ measurements after the change in emission rate which are easily detected by the changepoint detection algorithm. Moreover, above a leak rate ratio of 3, the changepoints are rarely missed if we account for delayed detection, therefore showing the effectiveness of the algorithm in raising the alarm when a substantial change in the emission rate occurs. The significance of this result can be highlighted by noting that in a recent study of natural gas well pads in California, it was shown that well pads for which facility-based emission estimates were at least 3 times the component-based emission estimations were responsible for 80\% the total measured emissions \cite{zhou_mobile_2021}. 

\begin{figure}
\centering
  \includegraphics[scale=0.65]{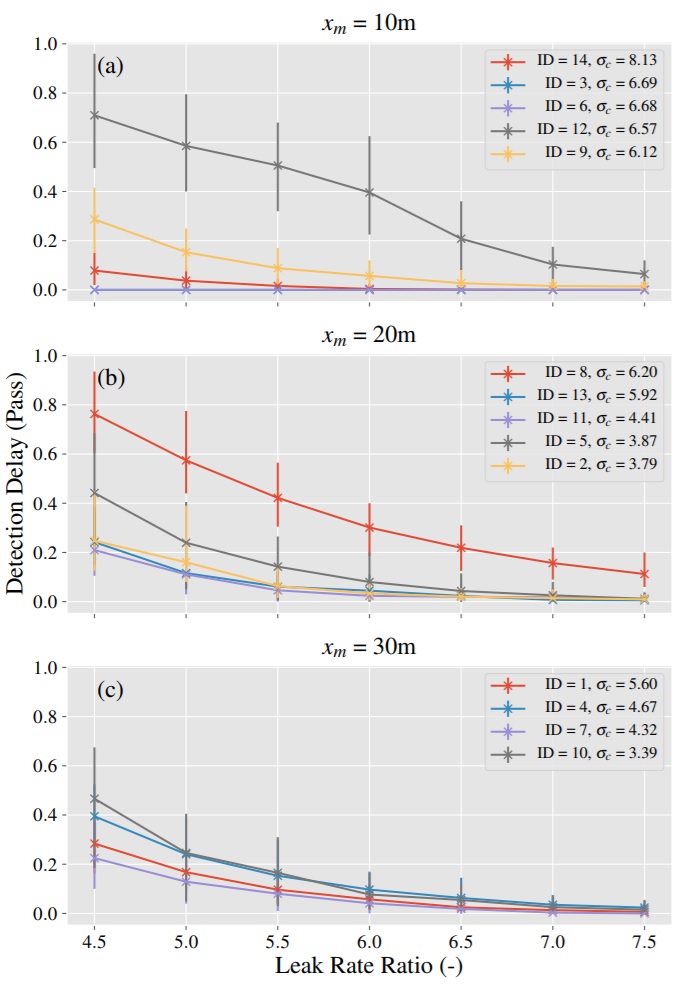}
  \caption{Evolution of detection delay (with units of number of passes) for a series of leak rate ratios varying between 4.5 and 7.5 for source-to-sensor distances, $x_m$ of (a) 10m, (b) 20m and (c) 30m. ID refers to the experiment ID as seen in Table \ref{Table1}, and $\sigma_c$ refers to the standard deviation of $c^y$ measurements before the changepoint that is calculated for each experiment. Vertical bars represent the 95\% confidence intervals.}
  \label{fgr_DetDel}
\end{figure}

Given that in some applications the change in emission rate can be intermittent, it is also important to quantify the delay in changepoint detection. Therefore, Figure \ref{fgr_DetDel} illustrates the detection delay for experiments where the changepoints are successfully detected across all instances (i.e., Detection Recall = 1) against increasing leak rate ratio. 
In this case, there is no clear trend between the noise in the measurements and the detection delay. However, the delay in changepoint detection monotonically decreases with increasing LRR as expected. It is worth noting that even at the lowest LRR where all changepoints are detected, the detection delay is less than one sensor pass, showcasing the speed of the changepoint detection algorithm.

\begin{figure}
\centering
  \includegraphics[scale=0.65]{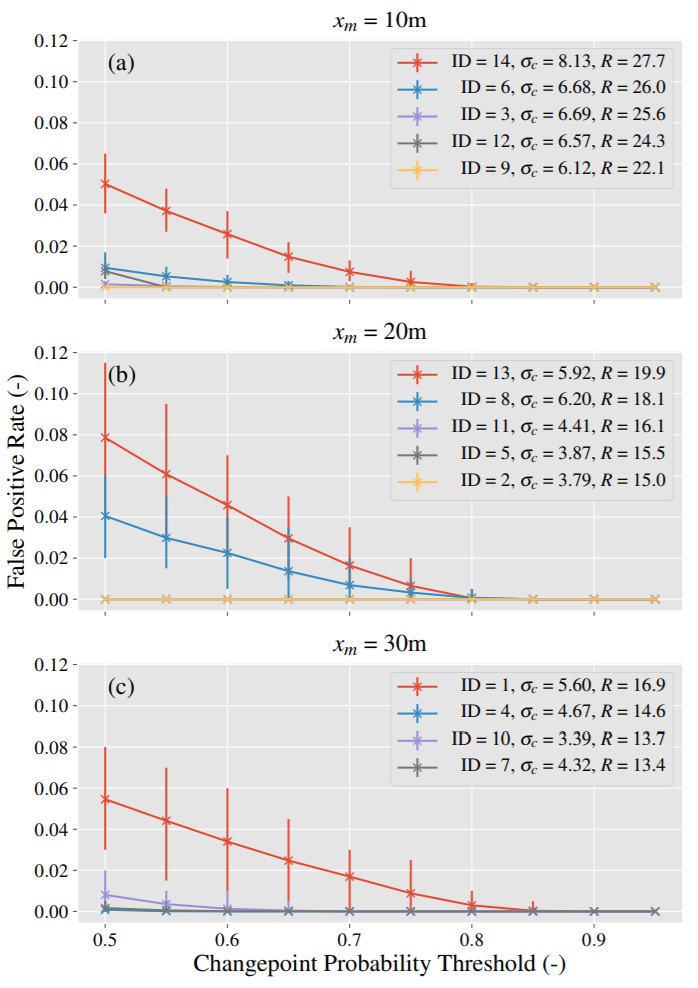}
  \caption{Evolution of false positive rate for changepoint probability thresholds varying between 0.5 and 0.95 for source-to-sensor distances, $x_m$ of (a) 10m, (b) 20m and (c) 30m. ID refers to the experiment ID as seen in Table \ref{Table1}, $\sigma_c$ refers to the standard deviation of $c^y$ measurements before the changepoint, and $R$ refers to range of $c^y$ measurements before the changepoint for each experiment. Vertical bars represent the 95\% confidence intervals.}
  \label{fgr_FP}
\end{figure}

Next, we investigate the sensitivity of the changepoint detection algorithm to the changepoint probability threshold, which in earlier results was set to a value of 0.8. Figure \ref{fgr_FP} presents the false positive rate when varying the changepoint probability threshold from 0.5 to 0.95. It can be seen that even at the lowest chosen threshold the false positive rate is less than 12\%, highlighting  the robustness of the changepoint detection algorithm. Furthermore, while the probability of false alarms is generally higher for experiments with larger noise, noise is not the sole predictor of the false positive rate in experiments. Range of the measurement distribution in each experiment, i.e., the difference between the maximum and minimum $c^y$ measurements in each experiment, seems to be a better predictor than standard deviation of measurements for the false positive rate. Consequently, the changepoint detection algorithm can be adversely affected by the presence of outliers in the data. There are multiple possible solutions for alleviating the sensitivity of the changepoint detection algorithm to outliers. One possible solution is to modify the changepoint detection condition first introduced in section \ref{sec:Fault}. For example, the condition can be adapted such that a changepoint is retained only if the changepoint probability is above a threshold for multiple measurements over the next few sensor passes. This requires the algorithm to delay resetting the Bayesian inference of equation (\ref{eq:Bayes}) until the detection condition is satisfied. The downside of this solution is potential poor changepoint detection when the change in emission rate is intermittent and temporary. 

\section{Conclusions}
In this study, we addressed the problem of detecting changes in the emission rate of a point-source by developing a recursive Bayesian scheme. This methodology directly builds on a recursive Bayesian framework that was previously used to estimate the emission rate from point sources. As a result, the introduced recursive Bayesian methodology has the ability to simultaneously detect changepoints in the emission rate and estimate the emission rate before and after changepoints. In addition, we applied our changepoint detection algorithm to a series of controlled release experiments, where a mobile sensor traversed cross-sections of the plume emitting from a point-source at different downwind distances, in the presence of an obstacle close to the source. Several measures were used to evaluate the performance of the changepoint detection methodology noting that the importance of each performance measure depends on the practical application at hand. We found that the changepoint algorithm is extremely effective ($>$90\% success rate) in identifying changes when the emission rate is tripled. This level of success is significant given recent findings suggesting that majority of emissions from the oil and gas sector can be caused by abnormal operations that drastically increase the emission rate \cite{zhou_mobile_2021,zavala-araiza_super-emitters_2017}. Further, the results showed that the statistics of the cross-plume mass concentration measurements such as the mean, standard deviation and the range can be used as predictors of the performance of the changepoint detection algorithm. Particularly, it was found that at a given leak rate ratio, lower values of coefficient of variation correspond to higher recall values which translates to higher effectiveness of the algorithm in detecting changes immediately after they occur. Moreover, it was shown that the false positive rate of the changepoint detection algorithm was less than 2\% when using a prescribed changepoint probability threshold of 0.8 for all the controlled release experiments. 

Although the changepoint detection algorithm was applied to mobile sensor measurements in the near field, the methodology can be easily adapted for fenceline monitoring applications using networks of fixed sensors or far-field measurements using a single stationary sensor. In these examples, mass concentrations and meteorological conditions are often averaged over 30-minute periods. By treating each 30-minute interval similar to a single sensor pass in the experiments described in the current study, changes in the emission rate can be found using the detection algorithm. 

With the changepoint detection methodology presented here applied to synthesized data from a single emission source, future work will be focused on evaluating the performance of the algorithm under more real-world scenarios such as intermittent faulty operation, and multiple emission rates caused by various operating conditions. Moreover, for practical settings, it is necessary to investigate the training time required to learn all the baseline parameters for the Bayesian inference scheme, most importantly the range of values used in the prior and the uncertainty term ($\sigma_e$) in the likelihood function of equation \eqref{eq:LF} before a change occurs. More studies on fault detection using advance sensing and measurement technologies will be beneficial in effective and rapid identification of large emitters which can lead to significant reductions in methane emissions from the oil and gas industry. 

\section*{Declaration of competing interest}
The authors declare they have no actual or potential competing financial interests.
\section*{Acknowledgments}
This project was supported by David R. Atkinson Center for a Sustainable Future (ACSF) at Cornell University, and DOE ARPA-E's Methane Observation Networks with Innovative Technology to Obtain Reductions (MONITOR) program under grant DE-AR0000749.

\clearpage
\bibliographystyle{ieeetr}  
\bibliography{CP_Project_V2}

\end{document}